\documentstyle[prd,aps,epsfig]{revtex}
\begin{document}
\title{A Predictive Minimal Model for Neutrino Masses and Mixings} 
\author{Berthold Stech\footnote{E-mail: B.Stech@thphys.uni-heidelberg.de}}
\address{Institut f\"ur Theoretische Physik, Universit\"at 
Heidelberg, Philosophenweg 16, D-69120 Heidelberg, Germany}
\maketitle
\begin{abstract}
A model is considered in which the scale of the heavy singlet
neutrinos is a few orders of magnitude below the grand unification
scale and where right-handed vector bosons play still a negligible
role. In a basis with diagonal up-quark and Dirac-neutrino mass
matrices it is assumed that the heavy neutrino mass matrix
has only zero elements in its diagonal, in analogy to the light neutrino
mass matrix in the Zee model. Connecting then the remaining matrix
elements with the small parameter describing the
hierarchy of quark masses and mixings and
by assuming commutativity of the charged lepton with the down-quark
mass matrix, the calculation of all neutrino properties can be performed
in terms of the two mass differences relevant for atmospheric and solar
neutrino oscillations. CP-violation is directly related to CP-violation
in the quark sector.
\end{abstract}
\section{Introduction}
Recent experiments give strong evidence that
neutrinos oscillate and have
finite masses \cite{1},\cite{2}. These experiments provide the first
conclusive evidence for physics beyond the standard model. Good candidates
for a corresponding extension of the standard model are supersymmetric
or non-supersymmetric grand unified theories \cite{3}.
For recent summaries in which many models are discussed and for further
literature see \cite{6}. Detailed grand unification
models need many assumptions about numerous new particles
and their representations, Higgs potentials and couplings.
One obtains interesting relations, but it is often hard to see to what 
extent the many assumptions are really necessary and decisive
for the light neutrino sector. In this article I will
formulate a model in which the heavy neutrinos get masses
several orders of magnitude below the scale of grand unification.
The possibility exists that these masses are generated
dynamically in some analogy to the generation of light
neutrino masses in the Zee model \cite{4}. Only
few particle properties at the unification scale are then
directly relevant for the neutrino
problem. The minimal model obtained in this way
provides an example in which the significance
of the assumptions are obvious and the neutrino 
properties can be calculated in all details.  
A disadvantage compared to an explicit treatment of a specific grand
unified theory is that there are no immediate additional predictions
about new physics at the unification scale.

Oscillations involving 3 light neutrinos\footnote{In this article the
possibility of the existence of light sterile neutrinos is not discussed
even though they are needed in a general analysis, in particular,
if the finding of the LNSD group \cite{5} are included.} can be described
in terms of 6 parameters: 2 parameters for the difference between
the square of the neutrino masses and 4 parameters for the unitary
matrix $U$ which relates -- in a basis in which the charged lepton
mass matrix is diagonal -- the flavor eigenstates $\nu_e,
\nu_\mu,\nu_\tau$ with the mass eigenstates $\nu_1, \nu_2, \nu_3$
\begin{equation}
\label{1}
\nu_\alpha=\sum_iU_{\alpha i}\nu_i\end{equation}
The theoretical understanding of neutrino physics and of these
parameters in particular requires, however, the consideration
of many more -- a priori unknown -- quantities.
An appealing way to extent the standard model is to add to the
3 two-component fields
new two-component fields $\hat\nu_e, \hat\nu_\mu, \hat\nu_\tau$, which are  
singlets with respect to the standard model gauge group and acquire large
masses corresponding to the scale of the new physics involved.
The well-known see-saw mechanism then provides for the observed
smallness of the light neutrino masses \cite{6}.

There are then 3 mass matrices
to consider\footnote{A classification of
three-neutrino models dealing with these mass matrices is
provided in ref. \cite{7}. Ans\"atze related to the present approach
can be found in ref. \cite{8}. I will closely follow and extend
my previous publications \cite{9} on this subject.}: the Dirac neutrino
mass matrix $m^{Dirac}_\nu$ which
connects the old with the new two-component fields, the mass matrix
for the charged leptons $m_E$ and the mass matrix of the singlet
neutrinos $M_R$.

There is some freedom for choosing a basis for the 3
matrices: We may take the Dirac neutrino mass matrix diagonal and real
which leaves us with 3 parameters for this matrix. Below the scale
of new physics the charged lepton mass matrix can be taken
hermitian. After a proper adjustment of the phases of the lepton
fields this matrix contains 7 parameters. Finally, the symmetric
mass matrix $M_R$ can be transformed to have real diagonal elements
and contains, therefore, 9 parameters.

In the next sections we will restrict these 19 parameters using the ideas
mentioned above and by suggesting connections with
the known structure of the mass matrices of up and down quarks.
The see-saw formula
\begin{equation}\label{2}
m_\nu=-m^{Dirac}_\nu\cdot M^{-1}_R\cdot(m^{Dirac}_\nu)^T\end{equation}
together with the mass matrix for charged leptons then allows
to calculate all the properties of light neutrinos of interest
here.

\section{The Dirac neutrino and the up-quark mass matrices}

Below the scale of new physics we are free to choose a basis
in which the Dirac neutrino mass matrix $m_\nu^{Dirac}$
and the up-quark mass matrix $m_U$
are diagonal simultaneously. In grand unified theories (taking
low Higgs representations) one finds that  at the unification
scale  $m_U$ and $m_\nu^{Dirac}$ are closely related and could
even be equal there \cite{10}. We will assume, therefore, that at the mass
scale of the heavy neutrinos, which will
turn out to be much lower than the unification scale,
$m_\nu^{Dirac}$ and $m_U$ can
be diagonalized simultaneously. In section 4 we will show that this
mass scale which we denote by $M_0$ is of order $10^{11}$ GeV.

The observed hierarchical structure of charged fermion masses
and mixings allows to express the corresponding mass matrices in terms
of powers of a small quantity. Using for this parameter  
$\sigma=(m_c/m_t)^{1/2}$, the diagonal up-quark mass matrix
can be written  \cite{9}
\begin{eqnarray}\label{3}
&&m_U(M_0)=\left(\begin{array}{ccc}
\sigma^4& 0&0\\
0&\sigma^2&0\\
0&0&1\end{array}\right) m_t(M_0)\nonumber\\
&&\sigma=0.057\qquad.\end{eqnarray}
With the help of the renormalization group equations Eq. (\ref{3})
together with the down-quark  and charged lepton mass matrices
(see section 5) provides
values for the masses of up, charm and top quarks at the scale
of the vector boson $Z$ which agree with the known result \cite{11}
\begin{eqnarray}\label{4}
m_u(m_Z)=1.9\pm0.4\ {\rm MeV}, && m_c(m_Z)=0.61\pm0.05\ {\rm GeV},
\nonumber\\
 m_t(m_Z)&=&173\pm5\ {\rm GeV},\nonumber\\
m_d(m_Z)=3.4\pm0.6\ {\rm MeV}, && m_s(m_Z)=0.064\pm12\ {\rm GeV},
\nonumber\\
 m_b(m_Z)&=&2.90\pm0.04\ {\rm GeV}\end{eqnarray}
within the given error limits.
For $m_\nu^{Dirac}$ at the scale $M_0$ we can write therefore
\begin{equation}\label{5}
m_\nu^{Dirac}(M_0)=\left(\begin{array}{ccc}
y\sigma^4& 0&0\\
0&x\sigma^2&0\\
0&0&1\end{array}\right) m_t(M_0)\end{equation}
with $x=O(1), \ y=O(1)$. There is no point in taking
an additional $O(1)$ parameter for the 33 element.
As can be seen below this would only affect the scale parameter
$M_0$ which is a fit parameter in our model.

\section{The mass matrix for the heavy singlet \protect\\ neutrinos}

In the absence of a detailed theory describing new physics beyond the
standard model, one can only speculate about the mass matrix for
the heavy neutrinos. However, one can hope that the small parameter
$\sigma$ which governs the up and down quark mass matrices plays also
here an important role \cite{9}. Because of the self-coupling of the heavy
neutrinos the assignment of generation quantum numbers
to those fields \cite{12} gives additional restrictions for the powers
of $\sigma$ occurring in this matrix. As an Ansatz for $M_R$
I will use here a matrix with only zero elements in its diagonal (see also
\cite{abud}),
in analogy to the matrix for the light neutrinos obtained in the
Zee model \cite{4}. $M_R$ could have a similar origin, this time, however,
involving the ``right-handed'' fields in a right-left symmetric
theory such as $SO(10)$ or $E6$ at the grand unification scale $\gg M_0$.
The elements of a matrix of this form can easily be connected
with the elements of the mass matrix $m_\nu^{Dirac}$ (and $m_U$)
with the powers of $\sigma$ related to appropriate generation charges
of the singlet neutrino fields. Therefore, I propose
\begin{equation}\label{6}
M_R=\left(\begin{array}{ccc}
0&x\sigma^3&\sigma\\ x\sigma^3& 0&a\\
\sigma& a& 0\end{array}\right)M_0\end{equation}
with $a=O(1)$. In this Ansatz, the ratio between $(M_R)_{12}$ and
$(M_R)_{13}$, namely $x\sigma^2$, has been taken to be identical
to the ratio between the elements $(m_\nu^{Dirac})_{22}$ and $(m_\nu
^{Dirac})_{33}$ in order to have a close relation with
$m_\nu^{Dirac}$. Since this ratio is a real number, all elements of
$M_R$ can be taken to be real by a proper phase choice for  the
singlet neutrino fields. The powers of $\sigma$ occurring in (\ref{6})
correspond to generation charges for the fields $\hat\nu_e,
\hat\nu_\mu,\hat\nu_\tau$ equal to $5/2, 3/2,-1/2$, respectively,
when setting the generation charge of the scalar field which produces
the heavy masses equal to $-1$.

$M_R$ can be diagonalized by an orthogonal matrix $V_R$ defined
by
\begin{equation}\label{7}
M_R=V_RM_R^{diagonal}V^T_R \quad .\end{equation}
The two large eigenvalues of $M_R$ differ up to order
$\sigma^2$ only in sign (like the eigenvalues of a Dirac neutrino mass
matrix) ($\simeq \pm aM_0$)
while the third eigenvalue is smaller by the factor $\sigma^4$.
Because of the almost degeneracy of two eigenvalues it is interesting to
look at the up-quark mass matrix when taken in a basis in which
$M_R$ is diagonal. One finds -- up to order $\sigma^2$ -- for $x=1$
\begin{equation}\label{8}
V_Rm_UV_R^T=\left(\begin{array}{ccc}
\frac{\sigma^2}{2a^2}&-\frac{\sigma}{2a}&-\frac{\sigma}{2a}\\
-\frac{\sigma}{2a}&\frac{1}{2}+\frac{\sigma^2}{2}(1-\frac{1}{a^2})
&\frac{1}{2}-\frac{\sigma^2}{2}(1+\frac{1}{4a^2})\\
-\frac{\sigma}{2a}&\frac{1}{2}-\frac{\sigma^2}{2}(1+\frac{1}{4a^2})
&\frac{1}{2}+\frac{
\sigma^2}{2}\end{array}\right)m_t(M_0).\end{equation}
Obviously, to a very good approximation, this matrix is of the
``democratic'' form in the 2,3 sector.

\section{The see-saw neutrino mass matrix}

The see-saw neutrino mass matrix $m_\nu$ can now be obtained
from Eq. (\ref{2}) using (\ref{5}) and (\ref{6}).
But before, we rescale the parameters $a$ and $M_0$ according to
$a\to\frac{x}{y}a,\ M_0\to y M_0$. One then gets the same expression
for the light neutrino mass matrix $m_\nu$ as in the special
case in which $x$ and $y$ in (\ref{5}) and (\ref{6})
are equal to 1:
\begin{equation}\label{9}
m_\nu(M_0)=-\frac{\sigma^2}{2aM_0}\left(\begin{array}{ccc}
-a^2\sigma^2& a\sigma& a\sigma\\
a\sigma& -1&1\\
a\sigma&1&-1\end{array}\right) (m_t(M_0))^2\end{equation}
The basis chosen is still the basis in which $m_U$ and $m_\nu^{Dirac}$
are diagonal matrices. Thus, we are not yet in a basis in which the
charged lepton mass matrix is diagonal. We also have to use the
renormalization group equation to go from the scale $M_0$ down to
the scale of the $Z$ boson. But one can calculate from (\ref{9})
the eigenvalues of the light neutrinos at $M_0$. As we will see
below, the pattern of these eigenvalues will not change
on the way down.

\noindent
The eigenvalues are
\begin{eqnarray}\label{10}
&&m_1(M_0)=-\frac{(m_t(M_0))^2}{M_0}\left(\frac{\sigma^3}{\sqrt2}-\frac{a \sigma^4}
{4}\right)\nonumber\\
&&m_2(M_0)=\frac{(m_t(M_0))^2}{M_0}\left(\frac{\sigma^3}{\sqrt2}+\frac{a  
\sigma^4}{4}\right)\nonumber\\
&&m_3(M_0)=\frac{(m_t^2(M_0))^2}{M_0}\frac{1}{a}
\sigma^2\end{eqnarray}
One can now identify the mass differences $m^2_3-m^2_1\approx m^2_3-m^2_2$
with $\Delta m^2$ observed in atmospheric neutrino experiments.
The Super-Kamiokande result \cite{2} $\Delta m^2
\approx 3\cdot 10^{-3}({\rm eV})^2$ gives for $aM_0$:
\begin{equation}\label{11}
aM_0\approx 5\cdot 10^{11}\ {\rm GeV}\end{equation}
For the ratio of the difference of squared masses, one gets
\begin{equation}\label{12}
\delta=\frac{m^2_2-m_1^2}{m^2_3-m^2_2}=\frac{a^3\sigma^3}{\sqrt2}+
O(\sigma^5)\end{equation}
Thus, the mass difference $m^2_2-m_1^2$ relevant for solar neutrino
oscillations turns out to be
\begin{equation}\label{13}
m^2_2-m^2_1\approx 4\cdot 10^{-7}a^3\ ({\rm eV})^2.\end{equation}
Interestingly, we are left with the single parameter $a$ only.

As long as the off-diagonal elements of the charged lepton mass matrix
are not much larger in magnitude than the off-diagonal elements of
the down-quark mass matrix (see the next section), the diagonalization
of (\ref{9}) provides already an estimate for the neutrino mixing matrix
$U$. The orthogonal matrix $O_M$ which diagonalizes $m_\nu(M_0)$ is
\begin{equation}\label{14}
O_M=\left(\begin{array}{ccc}
\frac{1}{\sqrt2}-\frac{a\sigma}{8}&-\frac{1}{\sqrt2}
-\frac{a\sigma}{8}&0\\
\frac{1}{2}+\frac{a\sigma}{8\sqrt2}&\frac{1}{2}-\frac{a\sigma}
{8\sqrt2}&-\frac{1}{\sqrt2}\\
\frac{1}{2}+\frac{a\sigma}{8\sqrt2}&\frac{1}{2}-
\frac{a\sigma}{8\sqrt2}&\frac{1}{\sqrt2}\end{array}\right)
+O(\sigma^2)\end{equation}
Because $\sigma/8$ is a very small number, one has the remarkable
result that the neutrino mixing matrix is of the bimaximal type \cite{14}
and very little dependent on the mass difference of the 2 lightest
neutrinos in the range acceptable for solar neutrino oscillations
\cite{1,6}.

\section{The charged lepton and the down-quark mass matrices}

We know the eigenvalues of the charged
lepton mass matrix from experiment but not its form
in a basis in which the Dirac neutrino
and the up-quark mass matrices are diagonal.
The down-quark mass matrix on the other hand is known to a
good extent from the measured down-quark masses and the
Cabibbo-Kobayashi-Maskawa mixing angles. In grand unified
theories one obtains relations between $m_E$ and $m_D$ \cite{10},
\cite{15},
but the details depend on the Higgs representations and Yukawa
couplings. The well-known relation obtained in $SU(5)$ \cite{3}, \cite{16}
\begin{equation}\label{15}
m^T_E\approx m_D\end{equation}
cannot be exact because of the different mass ratios in the quark and
lepton sector. Below the scale $M_0$ both $m_E$ and $m_D$
can be taken to be hermitian matrices. In the model considered
here this should still be possible when reaching the scale $M_0$,
which is much smaller than the unification scale where right-handed
vector bosons are effective. Otherwise
we would have to introduce more structure and thereby more unknown
parameters. From the relations in grand unified theories such as (\ref{15})
one can then conclude that the mixing for charged leptons are similar
to the mixing of quarks, i.e. not very large\footnote{In this respect
the present approach differs decisively
from models which use asymmetric matrices
by proposing  for the quark sector
large ``right-handed'' mixings of physical relevance
correlated with a large mixing in the ``left-handed'' lepton sector \cite{17},
\cite{YY}.}. The
mixing matrix $U$ for neutrinos (at $M_0$) will then not much differ
from $O_M$ given in (\ref{14}).

To be more specific I will consider the hypothesis that at $M_0$, $m^T_E$
and $m_D$ can be diagonalized simultaneously. In other words,
in our model both matrices, if put into hermitian forms, commute as
do the matrices $m_\nu^{Dirac}$ and $m_U$ \cite{9}.
\begin{equation}\label{16}
[m^T_E,m_D]=0\quad .\end{equation}
On the one hand, this condition is weaker than (\ref{15}) since
the matrix elements of $m_E^T$ and $m_D$ do not need to be equal or nearly
equal. On the other hand, it is more strict since (\ref{16}) allows
a complete calculation of $m_E$ from the lepton masses and $m_D$
(or the CKM matrix).
Of course all this has to be done at $M_0$ and should then be 
scaled down by
the renormalization group equations \cite{19} to the weak 
scale, i.e. the
scale of the $Z$-boson. For $m_D(M_0)$ we choose the matrix
\begin{eqnarray}\label {17}
m_D(M_0)&=&\left(\begin{array}{ccc}
0.7\sigma^3& i1.54\sigma^2& -i\sigma^2\\
-i1.54\sigma^2&-\sigma/3& i0.8\sigma\\
i\sigma^2& -i0.8\sigma& 1\end{array}\right)m_b(M_0) ~~,  \end{eqnarray} 
which together 
with $m_U$ of Eq. (\ref{3}) leads to mass eigenvalues at 
the weak scale within 
the experimental uncertainties (\ref{4}). 
It also gives at $m_Z$
\begin{equation}\label{18}
|V_{ud}|=0.22,\quad |V_{cb}|=0.039,\quad |V_{ub}|=0.0032\quad .\end{equation}

The phases chosen in (\ref{17}) are such as to obtain ``maximal''
CP-violation \cite{15,9} in the quark sector which may or may not be
supported by ongoing experiments.

An important matrix depending on the
off-diagonal elements of $m_D$ and which is relevant for the amount
of CP violation in the quark sector, is the commutator \cite{18}
\begin{equation}\label{19}
[m_U,\ m_D]=\frac{1}{i}C_q(M_0)\quad .\end{equation}
The analog commutator for leptons is
\begin{equation}\label{20}
[m_\nu^{Dirac},\ m^T_E]=\frac{1}{i}C_\ell(M_0)\quad .\end{equation}
As an alternative
to (\ref{16}) one can require that the off-diagonal elements of $m_D$
and $m_E$ have the same origin, i.e. arise from the same Higgs field:
\begin{equation}\label{21}
C_\ell(M_0)=\frac{m_\tau(M_0)}{m_b(M_0)}C_q(M_0)\quad .\end{equation}

The condition (\ref{21}), with $x=1$, $m_D$ from (\ref{17}), 
and the known charged 
lepton masses allows another calculation of $m_E(M_0)$
\begin{eqnarray}\label{22}
m_E(M_0)&=&\left(\begin{array}{ccc}
-0.66\sigma^3& -i1.54\sigma^2& i\sigma^2\\
i1.54\sigma^2&-\sigma& -i0.8\sigma\\
-i\sigma^2& i0.8\sigma& 1\end{array}\right)m_\tau(M_0)
\end{eqnarray}
This matrix differs somewhat from the one obtained from (\ref{16})
but leads again to small charged lepton mixing angles.

It would be interesting if another basis-independent  relation
between the mass matrices would exist, one which involves
the lepton mass matrices only. Since $M_R$ has no diagonal elements
in our model one could speculatively assume
\begin{equation}\label{23}
[m_\nu^{Dirac},\ m^T_E]=\frac{1}{i}\frac{m_t(M_0)m_\tau(M_0)}{M_0}
\sigma M_R(M_0) \quad.\end{equation}
Also this relation could be used to calculate $m_E$. The result,
although somewhat different from the one obtained for $m_E$ in
(\ref{22}) fixes the phases of $m_E$ and leads again to charged lepton  
mixing angles of similar small magnitudes.

\section{The calculation of the neutrino masses and the neutrino
mixing angles}

At the scale $M_0$ the mass matrix for light neutrinos is taken
from Eq. (\ref{9}). After the diagonalization of the mass matrix for 
charged leptons at this scale
\begin{eqnarray}\label{24}
m_E(M_0)=U_E(M_0)m_E^{diagonal}(M_0)\ U^\dagger_E(M_0)~,
\end{eqnarray}
$m_\nu(M_0)$ has to be transformed accordingly. The new neutrino
mass matrix is then
\begin{equation}\label{25}
\tilde m_\nu(M_0)=(U_E(M_0))^Tm_\nu(M_0)\ U_E(M_0)\quad.\end{equation}
This new neutrino matrix has now complex elements and will lead
therefore to CP-violation effects in neutrino processes.
$\tilde m_\nu(M_0)$ can now be scaled down to the weak
scale by the corresponding renormalization group equation. The evolution
equation according to the standard model is \cite{13}:
\begin{eqnarray}\label{26}
(4\pi)^2\frac{d}{dt}\tilde m_V&=&(-3g^2_2+2\lambda)\tilde m_\nu\nonumber\\
&&+\frac{4}{v^2}{\rm Tr}[3m_Um^\dagger_U+3m_Dm^\dagger_D+
m_Em^\dagger_E]\tilde m_\nu\nonumber\\
&&-\frac{1}{v^2}(\tilde m_\nu m_Em^\dagger_E+(m_Em^\dagger_E)^T\tilde
m_\nu)\quad.\end{eqnarray}
$\lambda=\lambda(t)$ denotes the Higgs coupling constant related
to the Higgs mass according to $m^2_H=\lambda v^2$
with $v=246$ GeV. We take $m_H(m_Z)=150$ GeV for the numerical
calculation. Solving (\ref{26}) allows to obtain the neutrino mass
matrix $\tilde m_\nu$ at the scale of the standard model. The neutrino
mixing matrix $U=U(m_Z)$ can then be obtained by
diagonalizing the hermitian matrix $\tilde m_\nu^*\tilde m_\nu$:
\begin{equation}\label{27}
\tilde m_\nu^*(m_Z)\ \tilde m_\nu(m_Z)=U(\tilde m_\nu^*\tilde m_\nu)^{diagonal}
U^\dagger.\end{equation}
The diagonal matrix
\begin{equation}\label{28}
\tilde m_\nu^{diagonal}=U^T\tilde m_\nu (m_Z)U\end{equation}
then provides the (complex) neutrino mass eigenvalues. By
introducing the diagonal phase matrix $\phi$ which consists
of the phase factors of $\tilde m_\nu^{diagonal}$ divided
by 2,  $U$ can be redefined $U\to U\phi$. The phase factors in
(\ref{28}) cancel and the so obtained unitary matrix $U$
expresses the neutrino states $\nu_e, \nu_\mu,\nu_\tau$ by
the neutrino mass eigenstates according to Eq. (\ref{1}).

The change of the mass matrices $m_U,\ m_D,\ m_E$ and $\tilde m_\nu$
between $m_Z$ and $M_0$ depends, of course, on the renormalization
group equations. To find the expression (\ref{17}) for $m_D(M_0)$ from
the -- approximate -- knowledge of $m_D(m_Z)$ I used the renormalization
group matrix equations of the standard model \cite{19}. It leads
to a change of $|V_{cb}|$ by about 10\%. When using the renormalization
group equation of the minimal supersymmetry model \cite{19}, the
main difference to the standard model case lies in the overall
factors $m_b(M)$, $m_t(M)$, which can easily be adjusted.
For $\tilde m_\nu(M)$ the formula (\ref{26}) has to be replaced by
the corresponding supersymmetry formula given by Babu et al. \cite{13}.

\section{Results and discussion}

It is straightforward to calculate from 
(\ref{9}),(\ref{16}),(\ref{17}),(\ref{24})-(\ref{28}) 
the light neutrino masses and the unitary mixing
matrix $U$ in terms of the parameters $M_0$ and $a$. In a very
direct way these two parameters determine the 2 mass differences
in the 3 neutrino scenario.
For the evolution of the mass matrices $m_U, m_D, m_E$ and $\tilde m
_\nu$ as a function of the scale we applied the renormalization
group equations according
to the standard model and according to the minimal
supersymmetry model. It turned out that for fixed $M_0$ and $a$ there is
no noticeable change in the ratio of the
3 eigenvalues of $\tilde m_\nu$ when going from $M_0$ down to $m_Z$. Even
when the squared mass difference of the two lightest neutrinos is
taken to be very small their mass ratio remains unchanged. Since both
masses have opposite signs as seen in (\ref{10}), radiative corrections
are ineffective as proved in ref. \cite{20}.

Interestingly, it also
turned out that the neutrino mixing matrix $U$ is practically
independent of the scale parameter. Moreover, the model predicts
$U$ to depend only very little on the difference of the neutrino
masses. The reason is that in the mixing matrix (\ref{14}) the
parameter $a$ plays a minor role only. Because also the
mixing angles of the charged leptons are small as described in
section 5, the matrix $U$ of our model is of the bimaximal
form \cite{14} for all mass differences of interest in neutrino oscillation
experiments\footnote{One may note here that for a
small value of $a$, $a\approx \sigma$, which describes a mass
squared difference between the lightest neutrinos of order $10^{-10}
({\rm eV})^2$, $m_\nu$ in (\ref{9}) takes a form not much different
from the neutrino mass matrix suggested in \cite{9} and gives
similar results.}.

The deviations from bimaximal mixing are therefore almost exclusively due to  
the mixings of charged leptons in the basis in which $m_\nu^{Dirac}$ is
diagonal. Using for $m_E(M_0)$ the result obtained from 
(\ref{16}) and (\ref{17}), one gets
\begin{equation}\label{29}
Abs[U(m_Z)]=\left(\begin{array}{ccc}
0.69& 0.71& 0.15\\
0.51& 0.51& 0.69\\
0.51& 0.49& 0.71\end{array}\right)\end{equation}
i.e. nearly pure bimaximal mixing.
The amount of CP violation depends evidently on the corresponding
CP-violating phase in the quark sector. With the phase choice taken
in (\ref{17}), which at the weak scale provides an acceptable form of the
conventional unitarity triangle for quarks, one obtains a neutrino
unitarity triangle with angles:
\begin{equation}\label{30}
\alpha_\nu\approx 77^o,\quad \beta_\nu\approx 17^o,\quad \gamma_\nu
\approx 86^o ~~.\end{equation} 
The form of this triangle is sensitive to the amount of CP 
violation in the
quark sector. When using instead of (\ref{16}) the condition (\ref{21}) 
i.e. $m_E(M_0)$ from (\ref{22}), (or from the condition (\ref{23}))
the triangle becomes more flat ~ 
($\alpha_\nu\approx 83^o,
\quad \beta_\nu\approx 6^o,\quad \gamma_\nu
\approx 91^o$)~ and $|U_{e,3}|\approx 0.06$.

In neutrino-less double $\beta$ decay experiments \cite{23} 
one measures the neutrino
mass averadge $|<\tilde m >_{e,e}|$. The model predicts the small
value
\begin{equation}\label{31}
 |<\tilde m >_{e,e}|~ \approx 0.002~ eV ~~.\end{equation}  

As we have seen, the model presented here is predictive . 
Essentially, only the two mass differences 
between the 3 neutrinos can be choosen freely. It predicts 
the bimaximal 
mixing form for the neutrino mixing matrix independently of 
the mass difference responsible for the solar neutrino 
oscillations. It is a minimal model because 
the parameter which describes the hierarchy in the 
quark sector could be used also for
the lepton  mass matrices, in particular, for the heavy neutrino
mass matrix, providing in this way for an intimate connection 
between quarks and 
leptons. It is a minimal model also
because only the simplest assumptions are made
for the connection between $m_\nu^{Dirac}$ and $m_U$ and between 
$m_E$ and $m_D$. The idea is that specific particles and interactions
acting at the grand unification scale, besides producing the heavy 
neutrinos at $M_0$, may play a minor role 
at this lower scale. To give $M_R$ the form of a Zee 
matrix is, of course, a strong assumption
but may seem not unreasonable for dynamically generated 
singlet neutrino masses 4 to 5 orders
of magnitudes below unification.
One obtains for the heavy neutrinos two almost degenerate 
mass eigenstates which are strongly mixed. Accordingly, in the 
special basis in which $M_R$ is diagonal, the matrices 
$m_U, m_D,  m_\nu^{Dirac}$ and $m_E$ are of a ``democratic'' 
form in the 2,3
sector (see Eq. (\ref{8})). In other words, in this specific
frame the Higgs field acts with (almost) equal strength 
on the members of the second and third family.
However, it is worth pointing out that one can relax the Zee ansatz
by including a 33 element to $M_R$. Because the inverse of $M_R$
enters the see-saw formula even a very large 33 element 
(of order $10^2$) still leads to near bimaximal mixing as long 
as the ratio $(M_R)_{12}/(M_R)_{13}$ used in (\ref{6}) 
is kept fixed. Thus,
the oscillation properties of the light neutrinos cannot distinguish
this strongly hierarchical mass matrix with small mixings of 
the heavy neutrinos\footnote{The possibility of having 
small mixings for the heavy neutrinos simultaneously with small 
mixings of the light leptons is discussed in \cite{XX}.}
from the mass matrix of Eq. (\ref{6}). 
\vspace{.3cm}

\noindent {\bf Acknowledgments}

\noindent The author likes to thank Dmitri Melikhov for a useful discussion.

\end{document}